\documentclass[PRD,reprint,onecolumn,linenumbers,amssymb,amsmath,aps,showpacs,nofootinbib,superscriptaddress,showkeys]{revtex4}
\usepackage{graphicx}
\usepackage{dcolumn}
\usepackage{bm}
\usepackage{color}
\usepackage{xspace}
\usepackage{ulem}

\newcommand{\fermi}{\emph{Fermi}\xspace}
\newcommand{\lat}{\emph{Fermi}-LAT\xspace}
\newcommand{\gr}{$\gamma$-ray\xspace}
\newcommand{\grs}{$\gamma$ rays\xspace}
\renewcommand{\deg}{\ensuremath{^{\circ}}\xspace}
\newcommand{\gev}{\ensuremath{\mathrm{GeV}}\xspace}
\newcommand{\mev}{\ensuremath{\mathrm{MeV}}\xspace}

\newcommand\fdg{\mbox{$.\!\!^\circ$}}
\newcommand{\src}{3FGL J2212.5+0703\xspace}
\newcommand{\etal}{et al.\xspace}

\newcommand{\logp}{{\tt LogParabola}\xspace}
\newcommand{\dmff}{{\tt DMFitFunction}\xspace}

\begin{document}

\title{
Testing the dark matter subhalo hypothesis of the gamma-ray source 3FGL J2212.5+0703
}

\author{Yuan-Peng Wang}
\affiliation{Key Laboratory of Dark Matter and Space Astronomy, Purple Mountain Observatory, Chinese Academy of Sciences, Nanjing 210008, China}
\affiliation{University of Chinese Academy of Sciences, Beijing, 100012, China}
\author{Kai-Kai Duan}
\affiliation{Key Laboratory of Dark Matter and Space Astronomy, Purple Mountain Observatory, Chinese Academy of Sciences, Nanjing 210008, China}
\affiliation{University of Chinese Academy of Sciences, Beijing, 100012, China}
\author{Peng-Xiong Ma}
\affiliation{Key Laboratory of Dark Matter and Space Astronomy, Purple Mountain Observatory, Chinese Academy of Sciences, Nanjing 210008, China}
\affiliation{University of Chinese Academy of Sciences, Beijing, 100012, China}
\author{Yun-Feng Liang$^\ast$}
\affiliation{Key Laboratory of Dark Matter and Space Astronomy, Purple Mountain Observatory, Chinese Academy of Sciences, Nanjing 210008, China}
\affiliation{University of Chinese Academy of Sciences, Beijing, 100012, China}
\author{Zhao-Qiang Shen$^\ast$}
\affiliation{Key Laboratory of Dark Matter and Space Astronomy, Purple Mountain Observatory, Chinese Academy of Sciences, Nanjing 210008, China}
\affiliation{University of Chinese Academy of Sciences, Beijing, 100012, China}
\author{Shang Li$^\ast$}
\affiliation{Key Laboratory of Dark Matter and Space Astronomy, Purple Mountain Observatory, Chinese Academy of Sciences, Nanjing 210008, China}
\affiliation{University of Chinese Academy of Sciences, Beijing, 100012, China}
\author{Chuan Yue}
\affiliation{Key Laboratory of Dark Matter and Space Astronomy, Purple Mountain Observatory, Chinese Academy of Sciences, Nanjing 210008, China}
\affiliation{University of Chinese Academy of Sciences, Beijing, 100012, China}
\author{Qiang Yuan}
\affiliation{Key Laboratory of Dark Matter and Space Astronomy, Purple Mountain Observatory, Chinese Academy of Sciences, Nanjing 210008, China}
\author{Jing-Jing Zang}
\affiliation{Key Laboratory of Dark Matter and Space Astronomy, Purple Mountain Observatory, Chinese Academy of Sciences, Nanjing 210008, China}
\author{Yi-Zhong Fan$^\ast$}
\affiliation{Key Laboratory of Dark Matter and Space Astronomy, Purple Mountain Observatory, Chinese Academy of Sciences, Nanjing 210008, China}
\author{Jin Chang}
\affiliation{Key Laboratory of Dark Matter and Space Astronomy, Purple Mountain Observatory, Chinese Academy of Sciences, Nanjing 210008, China}

\date{\today}

\begin{abstract}
N-body simulations predict that galaxies at the Milky Way scale host a large number of dark matter (DM) subhalos.
Some of these subhalos, if they are massive enough or close enough to the Earth, might be detectable in $\gamma$ rays due to the DM annihilation.
3FGL J2212.5+0703, an unidentified gamma-ray source, has been suggested to be the counterpart candidate of a DM subhalo by \citet{Bertoni2015,Bertoni2016}.
In this work we analyze the \emph{Fermi}-LAT Pass 8 data of 3FGL J2212.5+0703 to independently test the DM subhalo hypothesis of this source. In order to suppress the possible
contamination from two nearby very-bright blazars, we just take into account the front-converting gamma-rays which have better angular
resolutions than that of the back-converting photons. In addition to the spatial distribution analysis, we have extended the spectrum analysis down to
the energies of $\sim 100$ MeV, and thoroughly examined the variability of the emission during the past 8 years. We confirm that 3FGL J2212.5+0703
is a steady and spatially-extended gamma-ray emitter at a high confidence level. The spectrum is well consistent with that expected from DM annihilation into $b\bar{b}$.
The introduction of a phenomenological {\tt LogParabola} spectrum just improves the fit slightly.
All these results suggest that 3FGL J2212.5+0703 could be indicative of a DM subhalo.
\end{abstract}

\maketitle

\section{Introduction}
\label{sec_Intro}
Compelling evidence indicates that dark matter (DM) plays a significant role in many gravitational phenomena such as the galactic rotation curves, the galaxy cluster dynamics, and the cosmic microwave background \cite{Jungman1996, Bertone2005, Bertone2016}.
The latest measurements suggest that DM constitutes 84.3\% of the matter density in the current universe \cite{planck2016}.
Abundant as DM is, its particle physics nature remains unknown.
Various well-motivated DM candidates have been proposed in the literature and the leading candidate is the weakly interacting massive particles (WIMPs) \cite{Jungman1996, Bertone2005, Hooper2007, Feng2010, Bertone2016, Charles2016}.
WIMPs may annihilate or decay and finally produce
stable high-energy particle pairs, including for example electrons/positrons, protons/anti-protons, neutrinos/anti-neutrinos and gamma-rays.
These stable particles contribute to the cosmic radiation. The identification of these DM-originated particles in the gamma-ray and cosmic ray data is the prime goal of dark matter indirect detection. DM induced \grs are either from the decay or hadronization of the final state particles (i.e. prompt radiation), or from the final state particles interacting with interstellar medium or interstellar radiation field (i.e. secondary radiation). Unlike the charged particles that are deflected by the magnetic fields, gamma-rays travel straightforwardly and their morphology trace the distribution of the emitting sources directly. Therefore the searches for DM signal in the gamma-ray data, benefited from the spatial correlation with the DM distribution, have attracted wide attention. This is in particular the case after the successful launch of \emph{Fermi Gamma-ray Space Telescope} in June 2008 \cite{Atwood2009,Charles2016}.
Great efforts have been made to analyze the \lat (Large Area Telescope) data, but no reliable DM signal has been identified so far (see \cite{Charles2016} for a recent review).

Among various targets, the Milky Way dwarf spheroidal galaxies (dSphs), dominated by DM and in short of high energy astrophysical processes, are promising regions to identify the DM signal indirectly.
The identification of dSphs in optical however is rather challenging due to their low luminosities.
Until recently, only 25 dSphs were found by the Sloan Digital Sky Survey (SDSS) \cite{SDSS} and the observations prior to it (see \cite{McConnachie2012} and the references therein).
Over the past two years, 23 more dSphs (including candidates) have been discovered due to new optical image surveys \cite{Bechtol2015, Drlica-Wagner2015a, Koposov2015, Kim2015a, Laevens2015a, Laevens2015b, Martin2015, Kim2015b} such as the Dark Energy Survey (DES) \cite{DES} and the Pan-STARRS1 3$\pi$ survey \cite{Pan-STARRS}, or due to the reanalysis of the SDSS data. Although many analyses have been conducted in search for the \gr emission from these dSph sources and candidates \cite{Abdo2010, Ackermann2011, Geringer-Sameth2011, Tsai2013, Mazziotta2012, Cholis2012, Ackermann2014, Geringer-Sameth2015a, Geringer-Sameth2015b, Drlica-Wagner2015b, Ackermann2015, Hooper2015, Baring2016, Li2016, Ahnen2016, Liang2016, Albert2016}, none of them displays a statistically-significant signal (Tentative gamma-ray emission signals were reported in Ret II \cite{Geringer-Sameth2015b,Drlica-Wagner2015b} and Tuc III \cite{Li2016, Albert2016}).

DM subhalos are also promising targets for DM indirect detection.
In the standard hierarchical structure formation theory, dark matter particles accumulate to become small halos, and then merge repeatedly to form larger halos.
Some of the halos, if survived from the tidal stripping and virialization, become subhalos of the main halo \cite{White1978, White1991}.
N-body simulations at the scale of Milky Way show much more subhalos than satellites observed at optical wavelength \cite{VL-II, Aquarius}, indicating the majority of them contain little stars or gas.
DM subhalos, either massive enough or close enough to the Earth, could be visible in the gamma-ray band \cite{Springel2008, Yuan2012}.
More specifically, if 40 GeV DM particles annihilate with a cross section near the latest upper limit, \lat might have recorded $\sim 10$ DM subhalos \cite{Kuhlen2008, Anderson2010, Buckley2010, Belikov2012, Bertoni2015, Schoonenberg2016}.
Due to the short of stars and gas, DM subhalos may be only detectable in gamma-rays and hence are members of the unidentified gamma-ray sources.
Dedicated efforts have been made to search for such objects \cite{Buckley2010, Mirabal2010, Belikov2012, Zechlin2012a, Ackermann2012, Mirabal2012, Zechlin2012b, Berlin2014, Bertoni2015, Schoonenberg2016, Bertoni2016, Mirabal2016, Hooper2016}.
For instance, by fitting spectral energy distribution (SED) with DM spectra, bright high latitude point sources are investigated in \cite{Buckley2010, Belikov2012, Bertoni2015, Bertoni2016, Schoonenberg2016}. Candidates are selected from the spectrally hard sources and their multi-wavelength counterparts are examined in \cite{Zechlin2012a, Zechlin2012b}.
Independent source candidates created with looser assumptions on the spectrum are studied in \cite{Ackermann2012}.
Machine learning algorithms are also used to classify un-associated sources, and outliers, which are probably DM subhalos, are selected \cite{Mirabal2010, Mirabal2012, Mirabal2016}.
One very attractive finding is the identification of a spatially-extended source, \src, among the unidentified sources \cite{Bertoni2015, Bertoni2016} in the \lat third source catalog (3FGL). In astrophysical scenarios it is rather hard to give rise to a spatially-extended steady source without any association in other wavelength. Therefore \src is an interesting DM subhalo candidate, as stressed in \cite{Bertoni2016}.

Independent analysis is thus necessary to check whether it is indeed the case. For such a purpose, in this work we re-analyze the spatial, temporal and spectral characters of \src. The difference between this work and
\cite{Bertoni2015, Bertoni2016} are the following: (1)
In order to effectively suppress the possible contamination from two nearby extremely-bright blazars (3FGL J2254.0+1608 and 3FGL J2232.5+1143),
we just take into account the front-converting gamma-rays that have angular
resolutions better than the back-converting photons; (2) We have extended the spectrum analysis down to
the energy $\sim 100$ MeV and special attention has been given to the possible improvement in the fit by introducing a phenomenological shape in comparison to the $b\bar{b}$ model. Note that if significant improvement is found, the astrophysical origin will be favored; (3) The variability of the emission in the past 8 years rather than just the first four years has been thoroughly examined to better test the stability.

\section{Observation and data analysis}
\label{sec_ObsAna}

\lat Pass 8 is the most recent iteration of the event-level analysis, which reduces the ghost events (thus leading to an increased effective area and a better point-spread function), extends the energy reach, and introduces new event type partitions \cite{pass8}.
In this work, we use the front-converting Pass 8 Source data set\footnote{\url{ftp://legacy.gsfc.nasa.gov/fermi/data/lat/weekly/photon/}} (irfs=P8R2\_SOURCE\_V6, evtype=1) and the up-to-date \fermi\ {\tt ScienceTools}\footnote{version v10r0p5, available at \url{http://fermi.gsfc.nasa.gov/ssc/data/analysis/software/}}.
We select the data from October 27, 2008 to June 15, 2016 (i.e. Mission Elapsed Time (MET) range 246823875-487641604), and restrict the energy range from 100 \mev to 500 \gev.
To avoid significant contamination from the Earth's albedo, photons with zenith angle greater than 90\deg are excluded.
Quality-filter cuts ({\tt DATA\_QUAL==1 \&\& LAT\_CONFIG==1}) are also applied so as to remove the events and time intervals while the instrument is not in science configuration, or when either bright solar flares or particle events occur.\footnote{\url{http://fermi.gsfc.nasa.gov/ssc/data/analysis/documentation/Cicerone/Cicerone_Data_Exploration/Data_preparation.html}}
We consider the photons within a $21\deg \times 21\deg$ box centered on the position ($\alpha_{2000}=333\fdg147$, $\delta_{2000}=7\fdg0598$ \cite{3FGL}) so as to include at least 95\% photons\footnote{\url{http://www.slac.stanford.edu/exp/glast/groups/canda/lat_Performance.htm}} of \src at 100 \mev.
In order to perform binned analyses, photons are divided into $210 \times 210$ spatial bins and 30 logarithmic energy bins.

With the help of the user-contributed script {\tt make3FGLxml.py}\footnote{\url{http://fermi.gsfc.nasa.gov/ssc/data/analysis/user/}}, all 3FGL sources within 25\deg from the target source \cite{3FGL} as well as the Galactic diffuse \grs emission model {\tt gll\_iem\_v06.fits} and the isotropic emission template for the front-converting Source data selection {\tt iso\_P8R2\_SOURCE\_V6\_FRONT\_v06.txt}\footnote{\url{http://fermi.gsfc.nasa.gov/ssc/data/access/lat/BackgroundModels.html}} \cite{IEM} are included in the fit.
We leave free the spectra of the sources within 10\deg around \src, and the normalizations of the two diffuse emission backgrounds.
Note that one of the brightest blazars in the \gr band, 3FGL J2254.0+1608 (3C 454.3), is at the edge of the region of interest (ROI), whose influence is too significant to be ignored. Therefore we leave its spectral parameters free, even though it is outside the $10\deg$ radius circle.
We convolve all these models with the instrument response functions (IRFs) using {\tt gtsrcmaps}, and then perform the fittings with the {\tt pyLikelihood} code utilizing the {\tt MINUIT} algorithm \cite{Minuit}. As recommended in Fermi Science Center\footnote{\url{http://fermi.gsfc.nasa.gov/ssc/data/analysis/LAT_caveats.html}}, we take into account the energy dispersion of all free sources except the two diffuse emission models.

To check whether there is any significant residual in the ROI, we utilize the {\tt gttsmap} tool to generate test statistic (TS) maps, which are shown in Fig.\ref{tsmaps}.
The TS is defined as $-2\,{\rm ln}(\mathcal{L}_{\rm max,0}/\mathcal{L}_{\rm max,1})$ \cite{Mattox1996}, where $\mathcal{L}_{\rm max,1}$ and $\mathcal{L}_{\rm max,0}$ are the best-fit likelihood values of the alternative hypothesis (i.e. include a point source) and null hypothesis (i.e. background only), respectively.
We make a $20\deg \times 20\deg$ TS map and find three new point sources within 10\deg distance from the ROI center with a TS value larger than 25.
We first add these new point sources into the model with power-law spectra and coarse locations derived from the TS map, then do a fitting with energy dispersion turned off.
Using that fitted model as an input, the positions of these point sources as well as the target source \src are optimized with {\tt gtfindsrc}.
With the best fit positions, we further fit the spectra with energy dispersion enabled and the results are summarized in Tab.\ref{tab:newpts}.

\begin{table}[!htbp]
\begin{tabular}{lccr}
\hline\hline
Source Name & R.A.(\deg)    & Decl.(\deg)   &    TS \\
\hline
\src (pts)  & 333.13        &  7.06         & 320.2 \\
newpt0      & 327.22        & -1.33         &  25.9 \\
newpt1      & 329.05        & -0.61         & 146.7 \\
newpt2      & 332.77        & -0.06         &  36.3 \\
\hline
\end{tabular}
\caption{The optimized positions and the corresponding TS values of the target source (modeled with a point-source template) and the newly added point sources within 10\deg from the region of interest (ROI) center.}
\label{tab:newpts}
\end{table}

Note that our spectral parameters of the center source are in agreement with the values listed in the 3FGL catalog (within the $2\sigma$ confidence level).
Moreover, there is almost no shift in position for \src.

\subsection{Spatial Extension}
\begin{figure}[!h]
\includegraphics[width=0.45\textwidth]{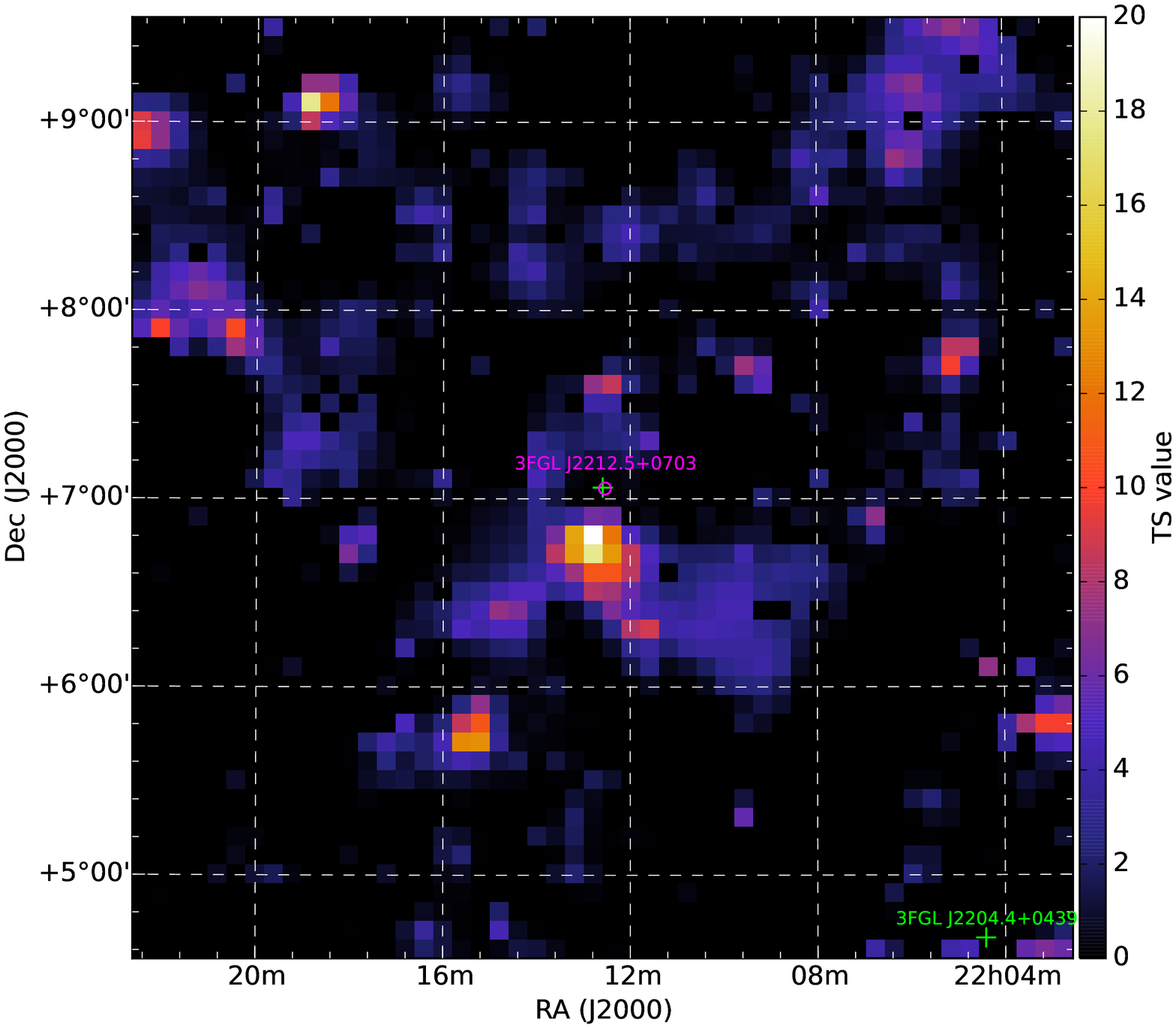}
\includegraphics[width=0.45\textwidth]{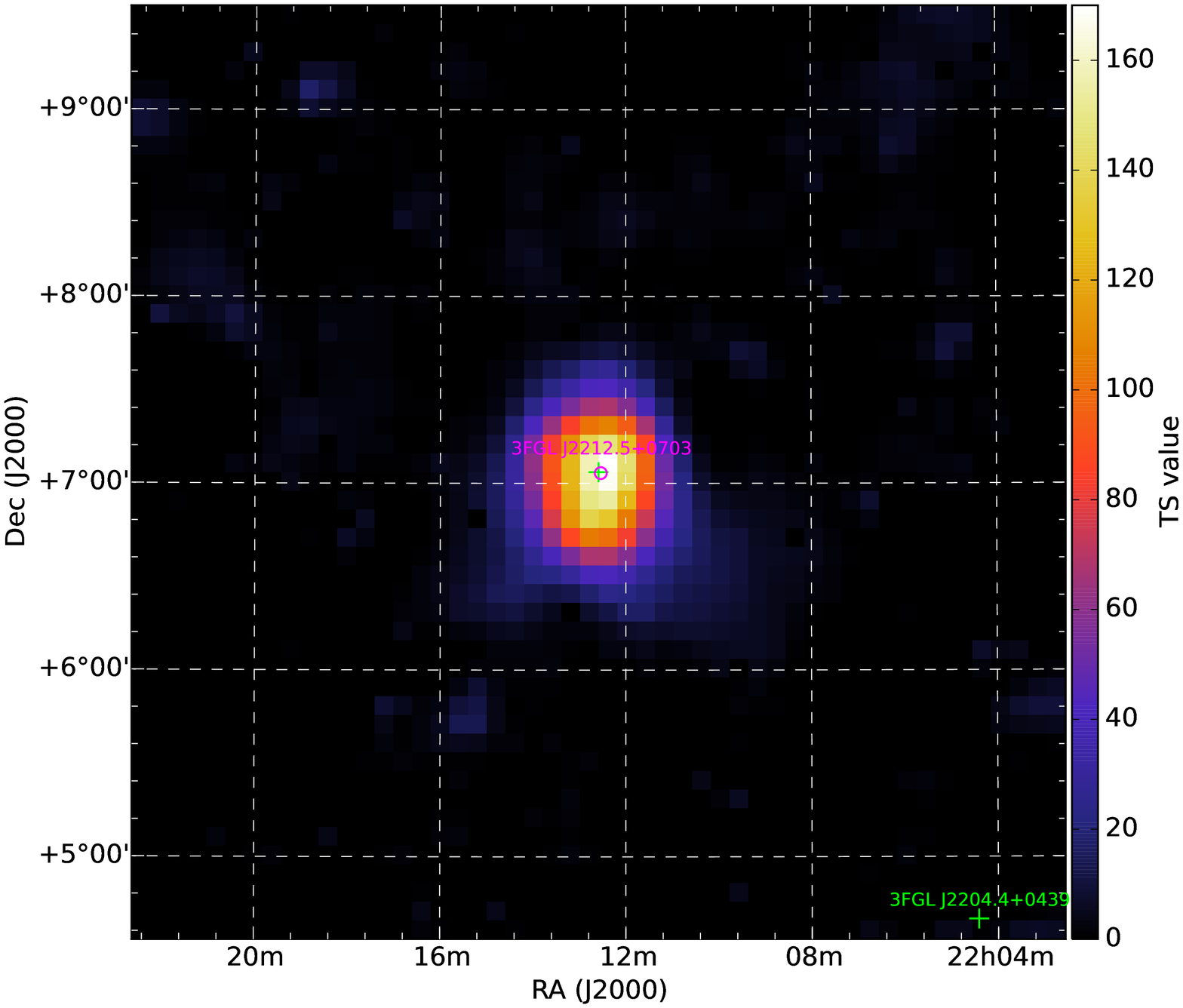}
\caption{
$5\deg \times 5\deg$ test statistic (TS) maps of the \src region using the front-converting Pass 8 data in the energy range of 1-500 \gev.
We show the TS map with (without) target source included in the model in the left (right) panel.
The green cross in the center of the figure is the position of target source listed in the catalog, while the magenta circle shows the position fitted with {\tt gtfindsrc}.
}
\label{tsmaps}
\end{figure}

Spatial extension is an important property for a source.
As argued in \cite{Bertoni2016}, an unambiguously spatially extended source without multi-wavelength associations could be a nearby DM subhalo.
In the left pannel of Fig.\ref{tsmaps}, some residuals remain near \src when we model it with a point source template.
To get a quick insight into the morphology of \src, we exclude it from the model and calculate the TS map again.
The TS map is exhibited in the right pannel of Fig.\ref{tsmaps}.
\src appears elliptical in TS map, and the residual in the left pannel of Fig.\ref{tsmaps} is caused by the incorrect spatial template.

\begin{figure}[!h]
\includegraphics[width=0.5\textwidth]{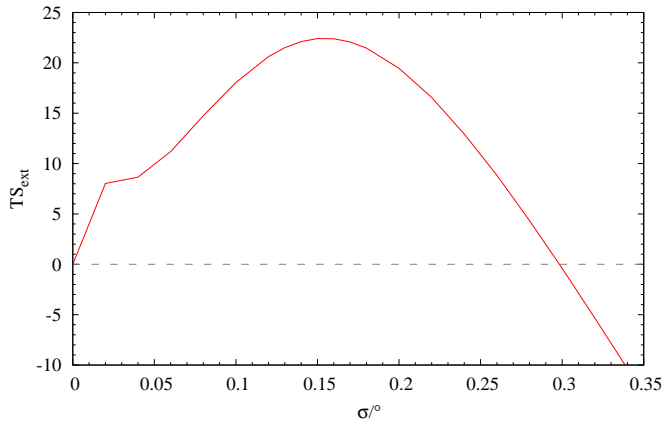}
\caption{
The increase in the TS value of \src when a Gaussian template instead of a point-like one is incorporated in the data fit.
The best fit template is the one with a width of 0\fdg15 and the corresponding ${\rm TS_{ext}}$ is 22.4.
}
\label{spatial}
\end{figure}

Following \cite{Bertoni2015}, to quantify the spatial extension we use a series of 2-D Gaussian distributions with different widths $\sigma$ as spatial templates.
Based on the optimized model in the previous subsection, we change the spatial template and perform optimizations.
The same likelihood ratio test is applied to achieve the best-fit spatial extension and the corresponding significance.
We define the TS for spatial extension to be
\begin{equation}
{\rm TS_{ext}}=-2\,{\rm ln} \left ( \frac{\mathcal{L}_{\rm point}}{\mathcal{L}_{\rm ext}} \right ),
\end{equation}
where $\mathcal{L}_{\rm point}$ and $\mathcal{L}_{\rm ext}$ are the best fit likelihood values for the point-source model and the Gaussian model, respectively \cite{Lande2012}.
Fig.\ref{spatial} depicts the relation between the widths and the TS values.
We find that the Gaussian template with a width of 0\fdg15 best describes the data and the corresponding ${\rm TS}_{\rm ext}$ value is 22.4, implying that the spatial-extended source model is better than the point-source model at a confidence level of $\sim 4.7 \sigma$ \cite{Chernoff1954,Lande2012}. Such an improvement is smaller than that reported in \cite{Bertoni2016} since in this work different spatial templates are used and just the front-converting Pass 8 data have been taken into account.

\subsection{The Light Curve and Variability}
The \gr signal from DM annihilation in a subhalo should be steady while the astrophysical signal may be variable.
With approximately 4 more years of data compared with 3FGL \cite{3FGL}, we can better test the compatibility of \src with a non-variable source.
The data are separated into 16 time bins equally.
We use the best-fit Gaussian template, and set free the normalization of the isotropic emission as well as the prefactors of the sources within 10\deg around \src.
To better model the nearby bright blazars, 3FGL J2254.0+1608 and 3FGL J2232.5+1143, we free both the normalizations and the indexes of them.
A fitting is performed in each time bin, and the likelihood profile of the target source is calculated.

\begin{figure}[!h]
\includegraphics[width=0.5\textwidth]{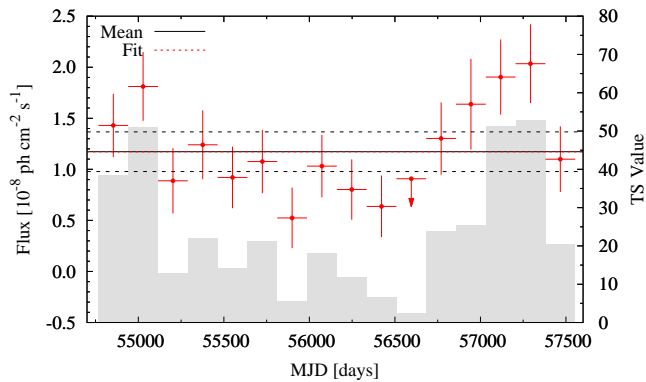}
\caption{
The light curve of \src.
Red points are the fluxes integrated from 100 \mev to 500 \gev, while the TS value of each time interval is shown in gray histogram.
95\% upper limits are drawn when the TS values are smaller than 5.
The solid line and two dashed horizontal lines represent the average flux and its 1$\sigma$ range, respectively.
The dotted line is the flux derived from the fit of the flux-likelihood relation in different time bins, which is almost the same as the average flux.
In this analysis, we find the variability index to be 33.2 with 15 degrees of freedom, indicating no significant ($\sim 2.6\sigma$) deviation from being a steady source.
}
\label{lc}
\end{figure}

The flux in each time bin is computed and the light curve of \src is yielded, as shown in Fig.\ref{lc}.
The variability index is defined as \cite{2FGL}
\begin{equation}
{\rm TS_{var}}=-2\,\sum_{\rm i} \frac{\Delta F_{\rm i}^2}{\Delta F_{\rm i}^2+f^2 F_{\rm const}^2} {\rm ln}\left( \frac{\mathcal{L}_{\rm i}(F_{\rm const})}{\mathcal{L}_{\rm i}(F_{\rm i})}\right ),
\end{equation}
where for the i-th time bin, $\mathcal{L}_{\rm i}$ is the likelihood value, $F_{\rm i}$ is the photon flux integrated over the energy range from 100 \mev to 500 \gev, $\Delta F_{\rm i}$ is the statistical uncertainty\footnote{We use the upper error of the flux.} of the flux $F_{\rm i}$. $F_{\rm const}$ is the flux if the source is constant, and $f$ is the systematic correction factor.
Following \cite{3FGL} we also take into account a 2\% systematic correction factor.
Using the likelihood profile we fit $F_{\rm const}$ in the above expression. The variability index is found to be 33.2 after the optimization, which corresponds to a significance of $\sim 2.6\sigma$ for the deviation from the non-variable hypothesis, for an approximate $\chi^2$ distribution with 15 degrees of freedom \cite{Wilks1938}.
The variability at time scales shorter than 6 months may not be effectively reflected in the above analysis, so we further do similar calculation using the data with one-month's time bin. To make the fitting stable, we fix the indexes of the two blazars mentioned above.
The variability index is found to be 123.2, corresponding to a $\sim 2.1\sigma$ significance for 92 degrees of freedom.

\subsection{Spectral Analysis}
\begin{figure}[!htbp]
\includegraphics[width=0.5\textwidth]{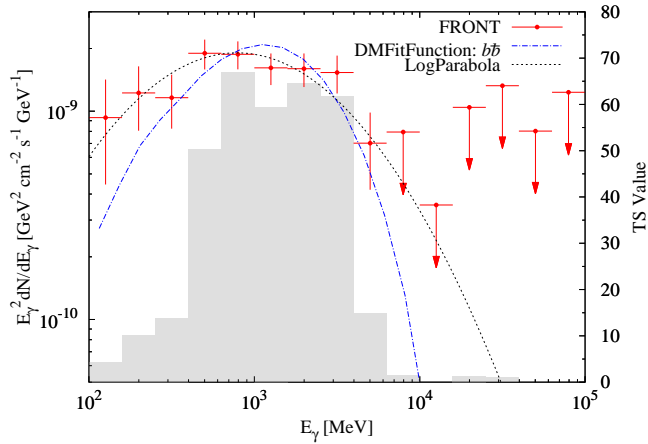}
\caption{
The spectral energy distribution (SED) of \src and two types of spectral modeling.
Gray histograms represent the TS values in different energy bins.
We use two types of spectra to fit the data, i.e. the generic \logp spectrum and the \dmff spectrum which predicts the annihilation of 18.5 \gev DM particles to $b\bar{b}$, and the corresponding spectra are shown in black (dashed) line and blue (dotted dashed) line, respectively.
The TS value difference between these two models is 7.0, implying that both models works almost equally well.}
\label{spectrum}
\end{figure}

The \grs from DM annihilation are characterized by a hard low-energy spectrum and a sharp cutoff at the mass of the DM particle,
so it is possible to distinguish between the DM model and the astrophysical origin model with the spectral information.
If an ``astrophysical" spectrum model can significantly better fit the gamma-ray data than the DM one, and meanwhile give a flat spectrum in low energy range, a DM origin would be disfavored. We compute the SED of \src to illustrate the spectrum in a model-independent way.
The data are binned with 15 equal logarithmic energy bins within the energy range from 100 \mev to 100 \gev.
We set the indexes of all sources frozen, and optimize their normalizations in each energy bin.
The intensity and TS value in each energy bin are shown in Fig.\ref{spectrum}.
According to the derived SED, we find that the spectrum of the target source is curved and can be well described by a
\logp spectrum, which is the phenomenological model given in the catalog. The DM spectrum model can well fit the data too (The spectrum might be a bit softer than the $b\bar{b}$ model at energies below 300 MeV. The error bars however are too large).

In order to compare the two spectral models quantitatively, we fit the data from 100 \mev to 500 \gev using the \dmff with the primary decay channel $b\bar{b}$ and the \logp model\footnote{We use the table {\tt gammamc\_dif.dat}, which is required in \dmff, contained in the {\tt FermiPy} package. {\tt FermiPy} can be downloaded from \url{https://github.com/fermiPy/fermipy}}, respectively.
The optimized spectra obtained in these fittings are also plotted in the Fig.\ref{spectrum}.
In the DM model, observational data favor a DM particle with a mass of $18.5~{\rm GeV}$.
Considering the probability function of DM mass may not be a Gaussian distribution, we make a likelihood profile to find a more accurate 1$\sigma$ confidence interval (CI) of DM mass.
We use a series of DM annihilation spectra from different DM masses, and do the fittings.
The edges of the 1$\sigma$ CI correspond to the models where the log-likelihood is 0.5 smaller than the maximum one \cite{Conrad2015}.
We find the 1$\sigma$ CI of DM mass to be (15.7, 20.9) \gev.\footnote{We also use the DM spectra from the PPPC4 \cite{PPPC4} and make the likelihood profile. We find the best fit mass is about 20.0 \gev and the 1$\sigma$ CI is (17.8, 23.2) \gev.}
We note the CI is different from that in \cite{Bertoni2016}, which is due to the different spatial morphology and analysis method.
The spectral TS,
\begin{equation}
{\rm TS_{spec}}=-2\, {\rm ln}\left( \frac{\mathcal{L}_{b\bar{b}}}{\mathcal{L}_{\rm logp}}\right ),
\end{equation}
is calculated, where $\mathcal{L}_{\rm logp}$ and $\mathcal{L}_{b\bar{b}}$ are maximum likelihood values for the \logp model and the \dmff model, respectively.
Through the fit, we find ${\rm TS_{spec}}$ to be 7.0, which indicates the \logp model is more compatible with the data.
Please bear in mind that the two spectral models are not nested in parameter space, so a significance can not be obtained from the Wilks theorem \cite{Wilks1938}. Since the TS value is quite small, the preference of the \logp model is weak.
More data are needed to draw a better conclusion.

\section{Discussion and conclusion}
\label{sec_discus}

DM subhalos are interesting targets for dark matter indirect detection. In principle some DM subhalos with little stars and dust may be only detectable in gamma-rays
and are thus members of unidentified gamma-ray sources. After the successful performance of \lat, great efforts have been made to identify such sources and
one very attractive finding is the identification of a spatially-extended source, \src, among the unidentified sources \cite{Bertoni2015, Bertoni2016} in the \lat 3FGL. Usually,
the astrophysical processes are hard to give rise to a spatially-extended steady source without any signals in other bands. Therefore \src is an interesting DM subhalo candidate,
as stressed in \cite{Bertoni2016}. Extended and independent analysis is necessary to check whether it is indeed the case. In this work we re-analyze the spatial, temporal and spectral characteristics of \src. In order to effectively reduce the possible contamination from two nearby extremely-bright blazars (3FGL J2254.0+1608 and 3FGL J2232.5+1143), just the
front-converting gamma-rays (with better angular resolutions than the back-converting photons) have been taken into account. With such data we confirm that \src is indeed a
spatially-extended source rather than a point-source and the optimized template has a width of 0\fdg15. We have also extended the spectrum analysis down to
the energy $\sim 100$ MeV and the main goal is to test whether an ``astrophysical" spectrum model (i.e., the phenomenological \logp model) can significantly better fit the gamma-ray data than the simple DM $b\bar{b}$ channel model. No significant improvement is found (The increase in ${\rm TS}$ is just $\sim 7$), implying that both the astrophysical model and the DM model (with a rest mass $m_\chi \sim 20$ GeV) work almost equally well. In principle, the spectral model can be better constrained as more data have been collected in the future. However, it is unlikely that the {\it Fermi}-LAT data for \src can be doubled. Even for the doubled gamma-ray data, it seems still challenging to distinguish between the two types of spectral models. Finally we checked the possible variability in the $\sim 8$ years' data and did not find significant evidence for deviation from a constant flux.

If \src is indeed a DM subhalo, its parameters can not be too extreme compared with those in N-body simulations.
Since the flux of the optimized DM model (with a rest mass $m_{\chi}=18.5~{\rm GeV}$) from 100 \mev to 500 \gev is $(8.2\pm0.6)\times10^{-9}~{\rm ph~cm^{-2}~s^{-1}}$, if the velocity-averaged annihilation cross-section of DM particles is assumed to be $\left<\sigma v\right>=2 \times 10^{-26}~{\rm cm^3~s^{-1}}$, we can derive that $\int_{V_{\rm sub}} \rho_{\rm sub}^2 (\bm{x}) d^3\bm{x} / (M_{\odot}^2~{\rm pc^{-3}}) = (4.7\pm0.3)\times 10^4~(D/{\rm kpc})^2$, where $\rho_{\rm sub}$ is the mass density of the subhalo and $D$ is the distance between the Sun and the source. Referring to clump luminosity-distance plane shown in Fig. 2 of \citet{Brun2009}, we find the target source is near the median distances calculated from a random sample of observer positions. Therefore a DM subhalo origin is not challenged.

Our conclusion is thus \src is indeed a steady spatially-extended un-identified gamma-ray source. These remarkable characteristics are compatible with the gamma-ray signal from a self-annihilating DM subhalo though how to yield such bright emission is still to be figured out. The other possibility that \src actually consists of two or more very nearby \gr sources also deserves a further investigation.
More \gr data as well as deeper multi-wavelength observations are needed to draw a final conclusion.

\begin{acknowledgments}
We would like to thank Dr. X. Li for useful discussions.
This research has made use of data obtained from the High Energy Astrophysics Science Archive Research Center (HEASARC), provided by NASA Goddard Space Flight Center.
This research has also used {\tt IPython} \cite{ipython}, {\tt NumPy}, {\tt SciPy} \cite{numpy}, {\tt Matplotlib} \cite{matplotlib}, {\tt Astropy} \cite{astropy}, {\tt F2PY} \cite{f2py}, {\tt APLpy}\footnote{\url{http://aplpy.github.com}} and {\tt iminuit}\footnote{\url{https://github.com/iminuit/iminuit}}.
This work was supported in part by the National Basic Research Program of China (No. 2013CB837000) and the National Key Program for Research and Development (No. 2016YFA0400200), the National Natural Science Foundation of China under grants No. 11525313 (that is the Funds for Distinguished Young Scholars), No. 11103084, No. 11303098 and No. 11303105, and the 100 Talents program of Chinese Academy of Sciences.
\end{acknowledgments}

$^\ast$Corresponding authors: liangyf@pmo.ac.cn (YFL), zqshen@pmo.ac.cn (ZQS), lishang@pmo.ac.cn (SL), yzfan@pmo.ac.cn (YZF).


\end{document}